\documentclass[prl,reprint,floatfix,letterpaper,twocolumn,nofootinbib,showpacs]{revtex4-1}

\usepackage[dvipdfmx]{graphicx}
\usepackage{bmpsize}
\usepackage{amsmath,amsfonts,amssymb}
\usepackage{hyperref}
\usepackage{braket}
\usepackage[caption=false]{subfig}
\hypersetup{
    bookmarksnumbered=true, %
    unicode=false, %
    pdfstartview={FitH}, %
    pdftitle={}, %
    pdfauthor={}, %
    pdfsubject={}, %
    pdfcreator={}, %
    pdfproducer={}, %
    pdfkeywords={}, %
    pdfnewwindow=true, %
    colorlinks=true, %
    linkcolor=blue, %
    citecolor=blue, %
    filecolor=blue, %
    urlcolor=blue %
}
\newcommand{\eq}[1]{(\hyperref[eq:#1]{\ref*{eq:#1}})}

\renewcommand{\sec}[1]{\hyperref[sec:#1]{Section~\ref*{sec:#1}}}
\newcommand{\thrm}[1]{\hyperref[thm:#1]{Theorem~\ref*{thm:#1}}}
\newcommand{\lemm}[1]{\hyperref[lemm:#1]{Lemma~\ref*{lemm:#1}}}
\newcommand{\prop}[1]{\hyperref[prop:#1]{Proposition~\ref*{prop:#1}}}
\newcommand{\corr}[1]{\hyperref[corr:#1]{Corollary~\ref*{corr:#1}}}
\newcommand{\fig}[1]{\hyperref[fig:#1]{Figure~\ref*{fig:#1}}}

\newcommand{\beginsupplement}{%
	\setcounter{table}{0}
	\renewcommand{\thetable}{S\arabic{table}}%
	\setcounter{figure}{0}
	\renewcommand{\thefigure}{S\arabic{figure}}%
}

\usepackage{xcolor}

\usepackage{tikz}

\usepackage{color} 
\usepackage{transparent}

\usepackage[caption=false]{subfig}
\usepackage{multirow}

\begin{document}

\title{Experimental demonstration of fault-tolerant state preparation\\ with superconducting qubits}
\author{Maika Takita}
\affiliation{IBM T.J. Watson Research Center, Yorktown Heights, NY 10598, USA}
\author{Andrew W. Cross}
\affiliation{IBM T.J. Watson Research Center, Yorktown Heights, NY 10598, USA}
\author{A. D. C\'orcoles}
\affiliation{IBM T.J. Watson Research Center, Yorktown Heights, NY 10598, USA}
\author{Jerry M. Chow}
\affiliation{IBM T.J. Watson Research Center, Yorktown Heights, NY 10598, USA}
\author{Jay M. Gambetta}
\affiliation{IBM T.J. Watson Research Center, Yorktown Heights, NY 10598, USA}

\begin{abstract}
Robust quantum computation requires encoding delicate quantum information into degrees of freedom that are hard for the environment to change. Quantum encodings have been demonstrated in many physical systems by observing and correcting storage errors, but applications require not just storing information; we must accurately compute even with faulty operations. The theory of fault-tolerant quantum computing illuminates a way forward by providing a foundation and collection of techniques for limiting the spread of errors. Here we implement one of the smallest quantum codes in a five-qubit superconducting transmon device and demonstrate fault-tolerant state preparation. We characterize the resulting  codewords through quantum process tomography and study the free evolution of the logical observables. Our results are consistent with fault-tolerant state preparation in a protected qubit subspace.
\end{abstract}

\maketitle

The possibility of robust quantum computation rests on the fact that quantum information can be encoded in degrees of freedom that are difficult for local noise processes to change. Quantum codes with this potential have been demonstrated in many physical systems \cite{Chow:2014,Corcoles:2015,Takita:2016,Cramer:2016,Riste:2015,Bell:2014,Nigg:2014,Reed:2012,Ofek:2016,Kelly:2015}. To make practical use of these codes, however, it is necessary not only to encode, decode, and observe errors, but to compute with faulty and inaccurate operations in a way that does not spread errors. The well-developed theory of fault-tolerant quantum computing reveals a steep experimental path toward this goal \cite{Gottesman:2010,Campbell:2017}. Recently, the question of what constitutes a {\it minimal} experimental demonstration of fault-tolerance was considered \cite{Gottesman:2016}. Fault-tolerant state preparation was demonstrated soon thereafter using a quantum error detecting code with trapped atomic ions \cite{Linke:2016}. Here we go beyond that result, implementing fault-tolerant state preparation on a superconducting qubit system with supporting evidence including quantum state tomography of prepared codewords, acceptance and logical error probabilities with and without error insertion, and analysis of the measured logical observables under free evolution.

We implement one of the smallest quantum codes, a four qubit code encoding two qubits \cite{Leung:1997}, and characterize output states produced by fault-tolerant state preparation circuits. The circuits are fault-tolerant for only one of the two encoded qubits, which allows direct comparison of their error rates. The circuits are applied in a five-qubit transmon device with nearest-neighbor connectivity. This device is a nontrivial subset of a surface code lattice in the sense that it provides resources for detection of any single-qubit error. Although the connectivity and size places limits on the set of fault-tolerant circuits we can implement on the four-qubit code, we can use stabilizer measurements to prepare codewords in a way that is analogous to surface code state preparation.

{\em Four qubit code} -- The four-qubit code \cite{Leung:1997} encodes two logical qubits into four physical qubits and can detect any error that acts on one of those physical qubits. It is the smallest code that can detect a general error and is unique \cite{Calderbank:1998}. The four-qubit code is defined by the stabilizer group $S = \langle S_x, S_z\rangle$ with stabilizers \cite{Gottesman:1997}
\begin{align}
S_x & = X_1X_2X_3X_4, \\ 
S_z & = Z_1Z_2Z_3Z_4. 
\end{align}
Here $X=|0\rangle\langle 1|+|1\rangle\langle 0|$ and $Z=|0\rangle\langle 0|-|1\rangle\langle 1|$ are Pauli operators.
The pair of encoded qubits are defined by logical operators
\begin{align}
\bar{X}_{L1} & = X_1X_3, & \bar{Z}_{L1} = & Z_1Z_2, \label{eq:logicalops} \\
\bar{X}_{L2} & = X_1X_2, & \bar{Z}_{L2} = & Z_1Z_3, \nonumber
\end{align}
The minimum distance of a stabilizer code is the minimum number of qubits acted on by any Pauli operator that commutes with $S$ but lies outside of it \cite{Gottesman:1997,Nielsen:2000}; in this case, that distance is two. Stabilizer codes are described by parameters $[[n,k,d]]$ where $n$ is the number of physical qubits, $k$ is the number of logical qubits, and $d$ is the minimum distance. Our code, thus, is a $[[4,2,2]]$ code.

The code space is spanned by four states
\begin{align}
|\bar{0}\bar{0},\tilde{0}\tilde{0}\rangle \propto |0000\rangle + |1111\rangle, \\
|\bar{0}\bar{1},\tilde{0}\tilde{0}\rangle \propto |1100\rangle + |0011\rangle, \\
|\bar{1}\bar{0},\tilde{0}\tilde{0}\rangle \propto |1010\rangle + |0101\rangle, \\
|\bar{1}\bar{1},\tilde{0}\tilde{0}\rangle \propto |0110\rangle + |1001\rangle.
\end{align}
On the left hand side, we order the labels $|L_1L_2,s_zs_x\rangle$ where $s_z$ and $s_x$ are syndrome bits that record phase and bit-flip errors, respectively. The syndromes correspond to single-shot measurements of the observables $S_x$ and $S_z$, which have eigenvalues $(-1)^{s_z}$ and $(-1)^{s_x}$, respectively.

We define destabilizers $\tilde{Z}_D = Z_4$ and
$\tilde{X}_D = X_4$ that commute with the logical operators and anticommute with corresponding stabilizers $S_x$ and $S_z$. The destabilizers change the values of the syndrome bits without affecting the logical qubits. The whole four-qubit Hilbert space is spanned by $16$ states $\{|L_1L_2,s_zs_x\rangle\}$ where $L_1$ and $L_2$ take values over the four states of the logical qubits and $s_z$ and $s_x$ run over the four possible syndromes. %

\begin{figure}
	\centering
	\includegraphics[width=3.375in]{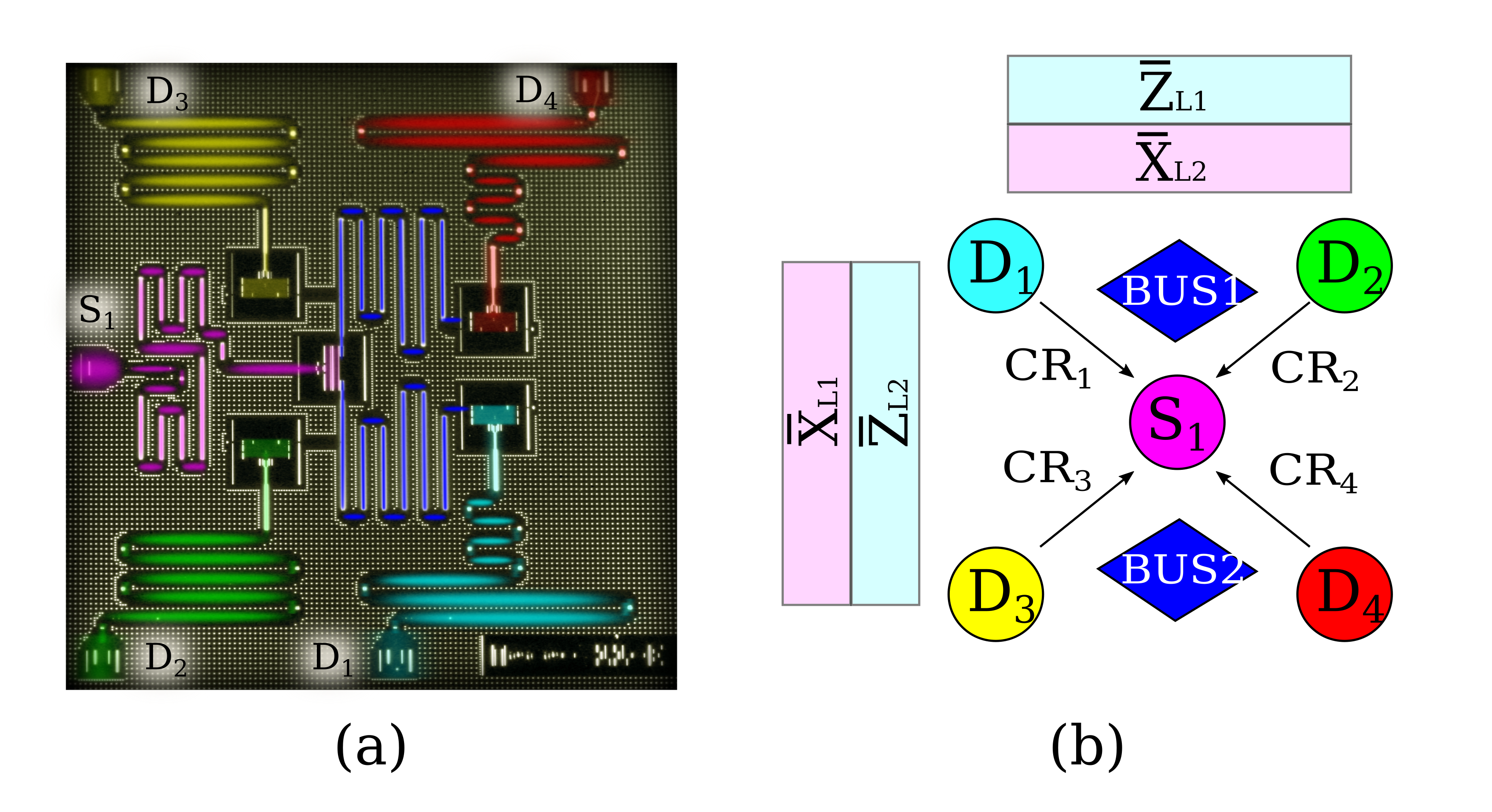}
	\caption{(color online) (a) False-colored micrograph of a five-qubit lattice. (b) Cartoon representation of a five-qubit lattice and the logical operators on the data qubits as given in Eq.~\ref{eq:logicalops}. Arrows represent the directions of the two-qubit cross resonance gate and point from the control to the target qubit.\label{figure:5Qfig} }
\end{figure}

{\it Implementation} -- The device consists of five fixed-frequency superconducting transmon qubits, four of which, $D_i$ with $i \in {1,2,3,4}$, are used as data qubits of the code (see Fig.~\ref{figure:5Qfig}). The central qubit, $S_1$, acts as a syndrome qubit, and it is coupled to the four data qubits via two coplanar waveguide (CPW) resonators acting as quantum buses, with two data qubits on each bus. Each qubit is coupled to its own CPW resonator for control and readout. Readout signals are amplified via Josephson Parametric Converters (JPCs)~\cite{Bergeal:2010,Abdo:2011aa}. Device fabrication methods are described in previous work \cite{Chow:2014,Corcoles:2015}. The device is the current IBM Quantum Experience device \cite{QX}.

Calibration training data is obtained from 4000 single shot measurements of $2^5 = 32$ different five-qubit computational states. Each measurement is a time-varying voltage signal reflected from the readout resonator. This signal is demodulated and integrated, yielding a single value in I/Q space. An arbitrary qubit state is determined by comparing the Euclidean distances between its integrated signal and the mean of the integrated signals of the ground and excited states obtained from the calibration data for that qubit. The shortest distance determines the outcome. The readout assignment error $\epsilon_r$ is given by

\begin{equation}
\epsilon_r = \frac{P(0|1)+P(1|0)}{2},
\end{equation}
where $P(\bar{b}|b)$ is the probability of observing the incorrect outcome given that the correct outcome is $b$. Readout assignment error for each qubit is given in the Supplemental Material.

Single-qubit gates are characterized using Clifford randomized benchmarking (RB)~\cite{Magesan:2011} and simultaneous RB~\cite{Gambetta:2012}. We find single qubit error per gate (EPG) of all five qubits to be lower than $\sim 9\times 10^{-4}$ and obtain crosstalk error of less than $\sim 6\times 10^{-4}$ from simultaneous RB results (see Supplemental Material for all measured single qubit errors). The two largest crosstalk errors are observed on D$_1$ and S$_1$, which is consistent with the fact that this pair of qubits has the largest static $ZZ$ interaction strength (see Supplemental Material).

\begin{figure}
	\includegraphics[width=3.2in]{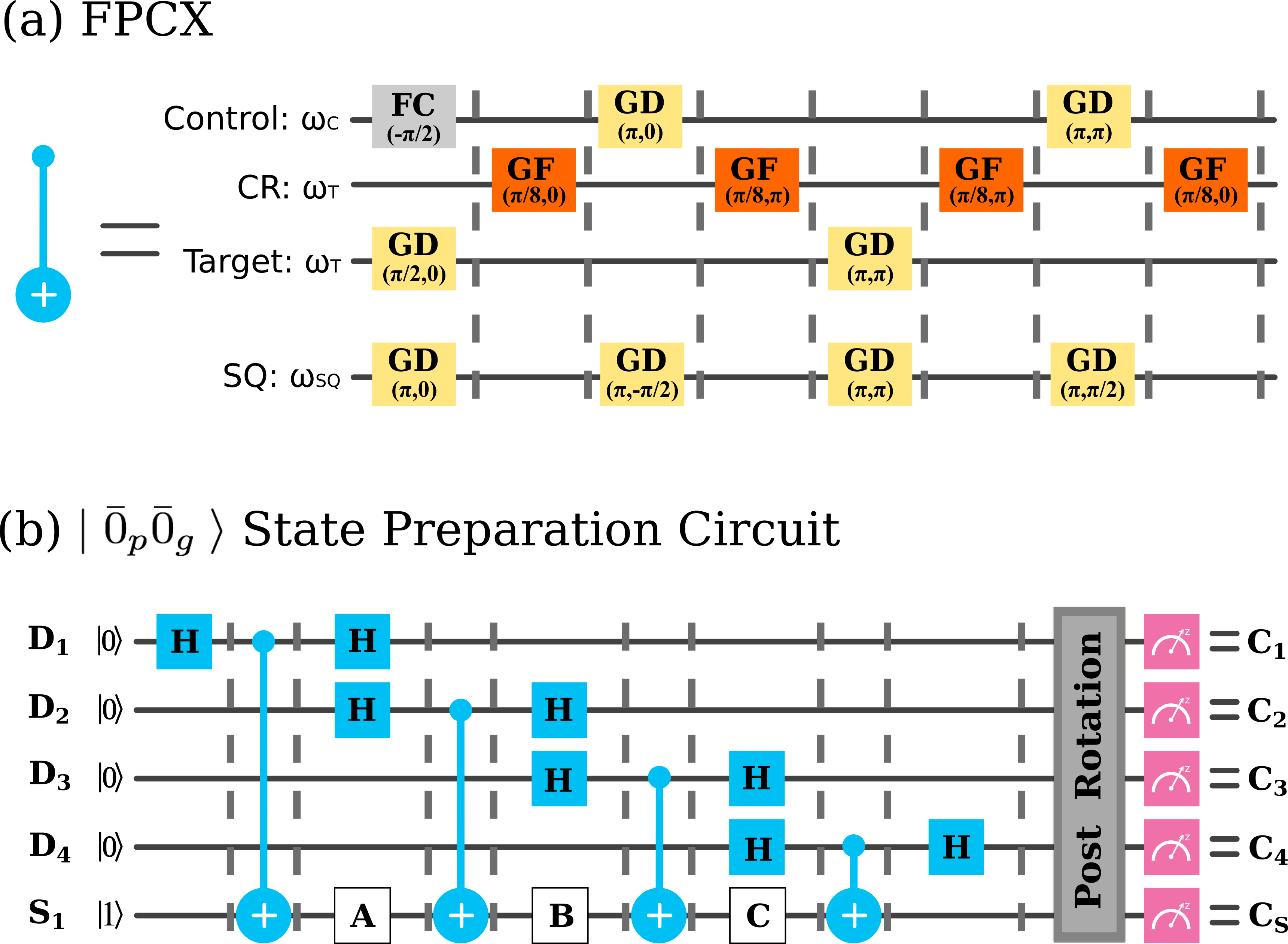}
	\caption{(color online) \textbf{CNOT pulse sequences and state preparation circuit.} (a) Decomposition of the four-pulse echoed CNOT gate (FPCX). Pulses are applied to the physical channels representing control, cross-resonance (CR), target, and spectator qubits (SQ).  The pulses comprise a frame change (FC) with an angle parameter, Gaussian derivative (GD) with an amplitude and angle, and a Gaussian flattop (GF) with an amplitude and angle. Spectator qubits are the three other qubits that are neither the control nor the target within the five-qubit lattice. (b) Logical state preparation circuit. $\ket{\bar{0}_p \bar{0}_g}$ is prepared without any post rotations (PR). Other states in logical Z-basis are prepared by applying $\bar{X}_{L1}$ and/or $\bar{X}_{L2}$. $\ket{\bar{+}_g \bar{+}_p}$ is prepared by applying the Hadamard gates on all four data qubits at PR. Other states in logical X-basis are prepared by applying $\bar{Z}_{L1}$ and/or $\bar{Z}_{L2}$ following the Hadamard gates at PR. Note that the first (left) logical qubit is the protected qubit in Z-basis but becomes the gauge qubit in X-basis due to the application of Hadamard gates at PR. \label{figure:Circuits}}
\end{figure}

Two-qubit controlled-NOT (CNOT) ~\cite{Paraoanu2006,Rigetti2010,Chow:2011} gates are constructed using the microwave-based cross resonance (CR) interaction. Using a four-pulse echoed cross resonance gate~\cite{Corcoles:2013,Takita:2016} as a two-qubit Clifford gate generator, we characterize four pairs of two-qubit gates through Clifford RB (Table~\ref{table:RB-result}). The decomposition of a four-pulse echoed cross resonance CNOT gate (FPCX) into single-qubit gates and CR interactions is drawn in Fig. ~\ref{figure:Circuits} (a). The FPCX echoes all of the first and second order \textit{Z}-terms from the cross-resonance Hamiltonian on the control, target, and the spectator qubits (SQ); the terms are \textit{ZII, IZI, IIZ, ZZI, ZIZ, IZZ}. Here, the spectator qubits are the three other qubits in the five-qubit lattice that are neither the control nor the target qubit for each particular CNOT. The FPCX sequence here is similar to the four pulse sequence used in previous work~\cite{Takita:2016} but has extra pulses to echo the \textit{IZI} term, which is typically smaller than the other terms. FPCX was necessary in order to correct for errors seen when using a two-pulse echoed cross resonance CNOT gate (TPCX) (See Supplemental Material for results using TPCX).

\begin{table}[ht]
	\begin{tabular*}{3.375in}{l|@{\extracolsep{\fill}}*{4}{c}}
		\hline \hline
		&$CR_1$ 	   &$CR_2$ 		&$CR_3$ 		&$CR_4$ 
		\\ \hline
		Two-qubit 		&0.0380  		&0.0451 	&0.0330 	&0.0282\\
		EPG	&$\pm$ 0.0013 	&$\pm$ 0.0015 	&$\pm$ 0.0012 	&$\pm$ 0.0010  \\ 
		(length, ns)	&(780) 		&(780)	&(780)	&(780) \\
		\hline 	\hline			
	\end{tabular*}
	\caption{\label{table:RB-result} \textbf{Two-qubit error per gate (EPG).} Randomized benchmarking results for the two-qubit gate with four cross-resonance pulses. The two-qubit error-per-gate (EPG) and total sequence length are shown.}
\end{table}

{\it Fault-tolerant state preparation} -- The logical state $\ket{\bar{0}_p \bar{0}_g}$ is prepared by running the X-stabilizer ($S_x$) circuit and measuring the syndrome qubit (shown in Fig.~\ref{figure:Circuits}). The logical qubit $L_1$, denoted here by \textit{p}, is a fault-tolerantly prepared protected qubit, and $L_2$, denoted by \textit{g}, is a gauge qubit that is not prepared fault-tolerantly. The other logical states in the $\ket{\bar{0}}$ and $\ket{\bar{1}}$ basis are prepared by applying logical bit-flips, Eq.~\ref{eq:logicalops}, during the $\ket{\bar{0}_p \bar{0}_g}$ state preparation. From the $\ket{\bar{0}_p \bar{0}_g}$ state, we can prepare $\ket{\bar{+}_g \bar{+}_p}$ by applying Hadamard gates on the four data qubits. Note that this swaps the indices of the logical states, exchanging $\bar{X}_{L1}$ with $\bar{Z}_{L2}$ and $\bar{X}_{L2}$ with $\bar{Z}_{L1}$. The other logical states in the $\ket{\bar{+}}$ and $\ket{\bar{-}}$ basis are prepared by applying logical phase-flips, Eq.~\ref{eq:logicalops}, after the Hadamard gates.

\begin{figure}
	\includegraphics[width=3in]{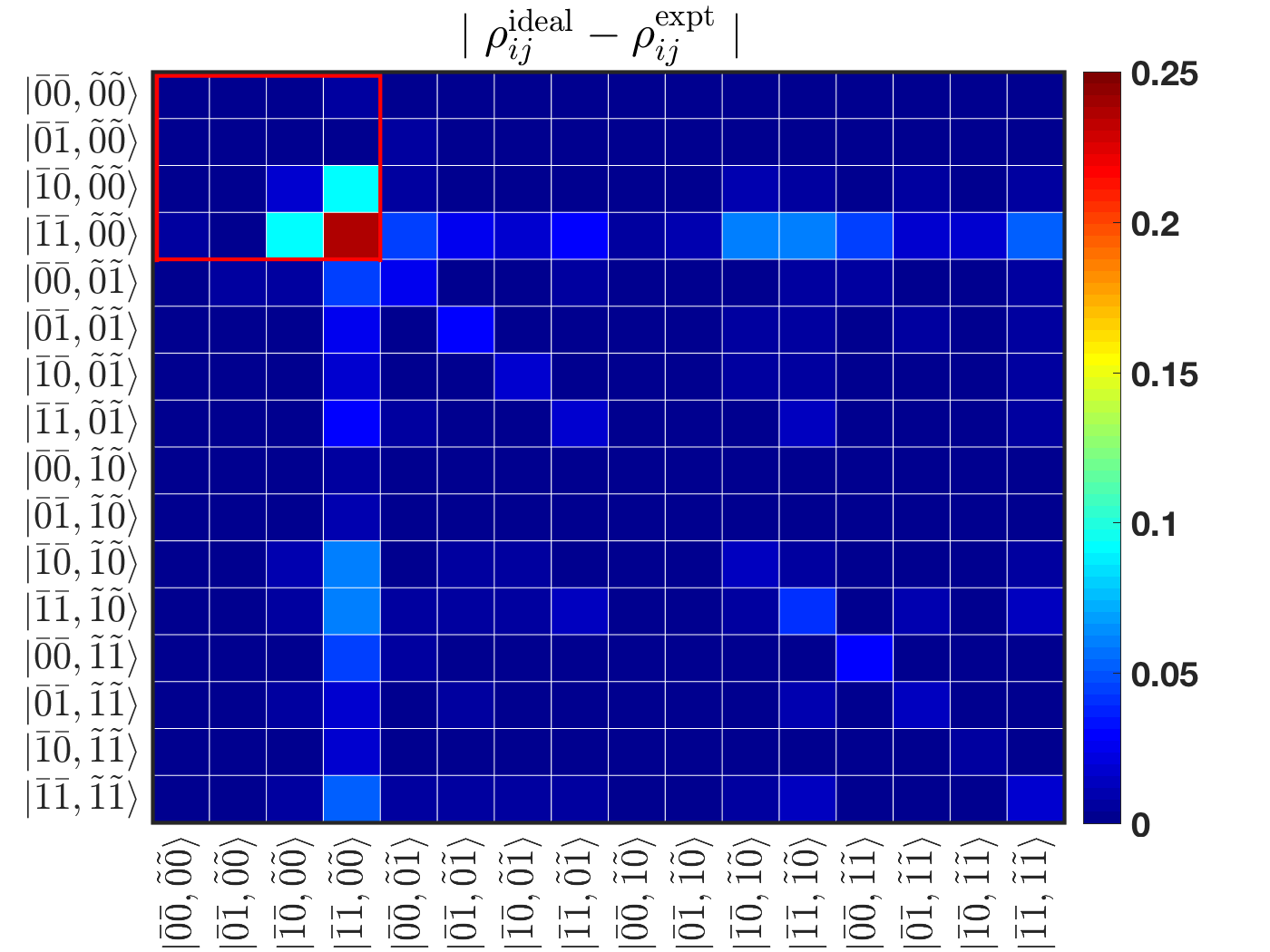}
	\caption{(color online) \textbf{Magnitude of the reconstructed $|\bar{1}_p\bar{1}_g\rangle$ state.} We show the absolute differences between the actual and ideal matrix elements in the basis consisting of $16$ states $\{|L_1L_2,s_zs_x\rangle\}$. Labels $L_1$ and $L_2$ run over the four states of the logical qubits in $Z$-basis. Syndrome bits $s_z$ and $s_x$ run over the four possible syndromes and represent the presence of phase-flip and bit-flip errors, respectively.\label{figure:state-tomo-FPCX}}
\end{figure}

To characterize the state preparation circuit, we performed quantum state tomography on the four data qubits. The difference between the ideal and reconstructed state of $\ket{\bar{1}_p \bar{1}_g}$ is shown in Fig.~\ref{figure:state-tomo-FPCX}. The boxed top left corner of the reconstructed state represents the projection onto the codespace $(\tilde{0}\tilde{0})$. Considering the corresponding state $\rho(\tilde{0}\tilde{0})$, the largest errors are coherent errors on the gauge qubit. The acceptance probability $\mathrm{tr}(\rho_{\tilde{0}\tilde{0}})$ and fidelity of the prepared state $\rho$ are obtained from $\rho_{\tilde{0}\tilde{0}}$. Results computed from state tomography data of additional prepared logical states are given in Table~\ref{table:FPCX-stateprep}.

\begin{table}[ht]
	\begin{tabular*}{3.375in}{l|@{\extracolsep{\fill}}*{5}{c}}
		\hline \hline
		Prepare&  Accept 	&$\ket{\bar{0}_p\bar{0}_g}$	&$\ket{\bar{0}_p\bar{1}_g}$	&$\ket{\bar{1}_p\bar{0}_g}$	&$\ket{\bar{1}_p\bar{1}_g}$ \\ \hline	
		$\ket{\bar{0}_p\bar{0}_g}$	&0.7566	&0.9726	&0.0216	&0.0040	&0.0019	\\ %
		$\ket{\bar{0}_p\bar{1}_g}$	&0.7773	&0.0245&0.9678	&0.0037	&0.0041	\\ %
		$\ket{\bar{1}_p\bar{0}_g}$	&0.7702	&0.0028	&0.0042	&0.9673	&0.0258	\\ %
		$\ket{\bar{1}_p\bar{1}_g}$	&0.7853	&0.0033	&0.0034	&0.0224	&0.9709	\\ %
		\hline \hline
		Prepare&  Accept 	&$\ket{\bar{+}_g\bar{+}_p}$	&$\ket{\bar{+}_g\bar{-}_p}$	&$\ket{\bar{-}_g\bar{+}_p}$	&$\ket{\bar{-}_g\bar{-}_p}$ \\ \hline	
		$\ket{\bar{+}_g\bar{+}_p}$	&0.7897	&0.9667	&0.0065	&0.0199	&0.0069	\\ %
		$\ket{\bar{+}_g\bar{-}_p}$	&0.7707	&0.0057	&0.9632	&0.0064	&0.0247	\\ %
		$\ket{\bar{-}_g\bar{+}_p}$	&0.7799	&0.0247	&0.0069	&0.9626	&0.0058	\\ %
		$\ket{\bar{-}_g\bar{-}_p}$	&0.7731	&0.0065	&0.0253	&0.0063	&0.9619	\\ %
		\hline \hline								
	\end{tabular*}
	\caption{\label{table:FPCX-stateprep} Acceptance probability and fidelity of initial state preparation given that the state is in the codespace, $s_zs_x=00$. The largest errors are due to error on the gauge qubit.}
\end{table}

{\it Error insertion} -- To study how error propagates through the $\ket{\bar{1}_p\bar{1}_g}$ state preparation circuit, we introduce a phase error $Z(\theta)$ on $S_1$ after the 1st (A), 2nd (B), or 3rd (C) CNOT gate [see Fig.~\ref{figure:Circuits} (b)]. Since the state preparation is done by syndrome measurement, we first post-select on the syndrome measurement reading c$_s= 1$, noting that the syndrome qubit starts from the excited state at the beginning of the circuit. Next, we compute $S_z$ in software and post-select on $\text{c}_1 \oplus \text{c}_2 \oplus \text{c}_3 \oplus \text{c}_4 = 0$. The acceptance probability is given by P($\text{c}_1 \oplus \text{c}_2 \oplus \text{c}_3 \oplus \text{c}_4 = 0 | \text{c}_s= 1$), and the state of the protected (gauge) qubit is determined from the parity of $\text{c}_1$ and $\text{c}_2$ ($\text{c}_1$ and $\text{c}_3$). 

\begin{figure}
	\subfloat[Acceptance $Z(\theta)$ \label{figure:Error-Accept}]{%
		\includegraphics[width=1.65in]{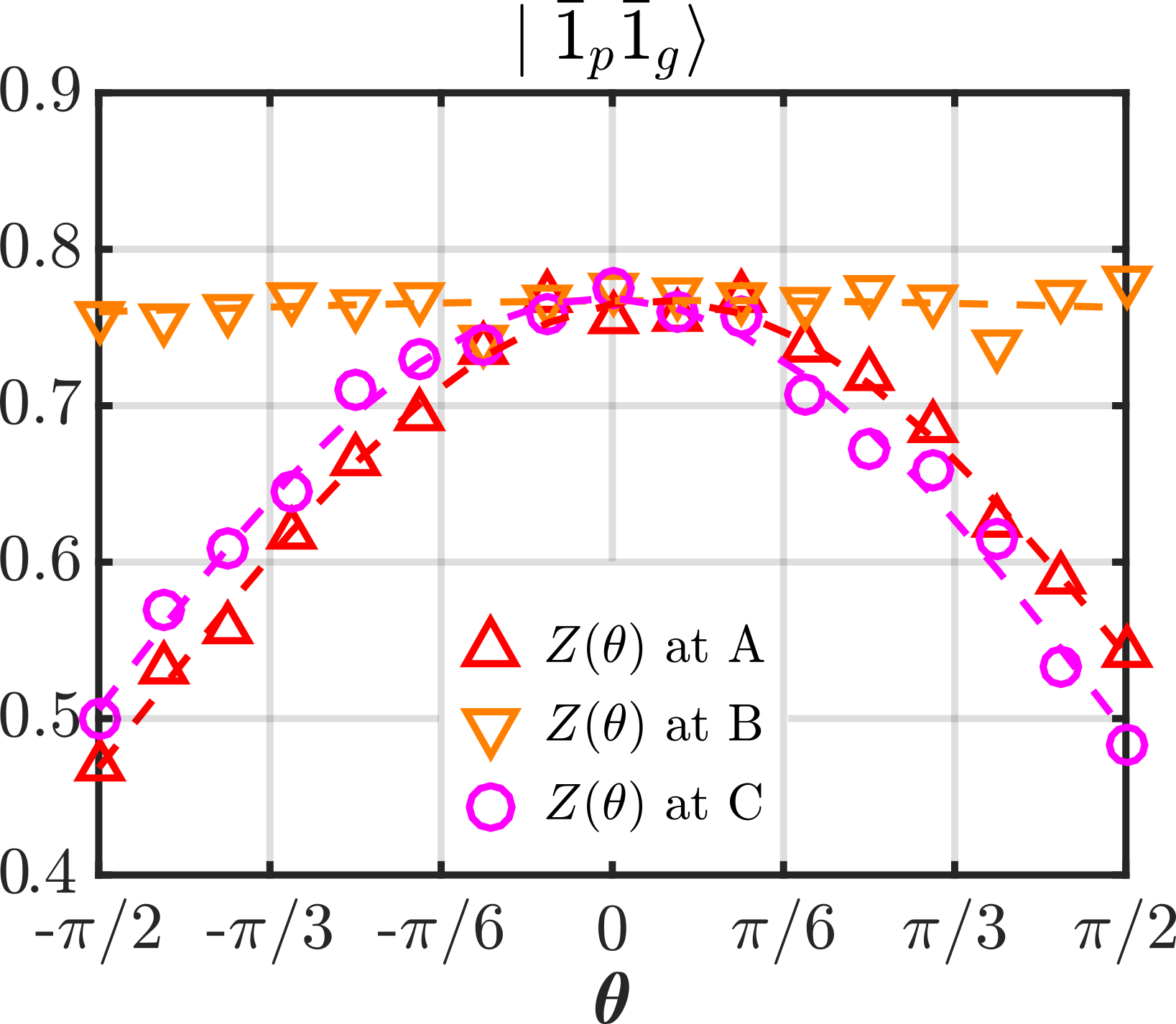}
	}
	\hfill	
	\subfloat[Error $Z(\theta)$ \label{figure:Error-probability}]{%
		\includegraphics[width=1.65in]{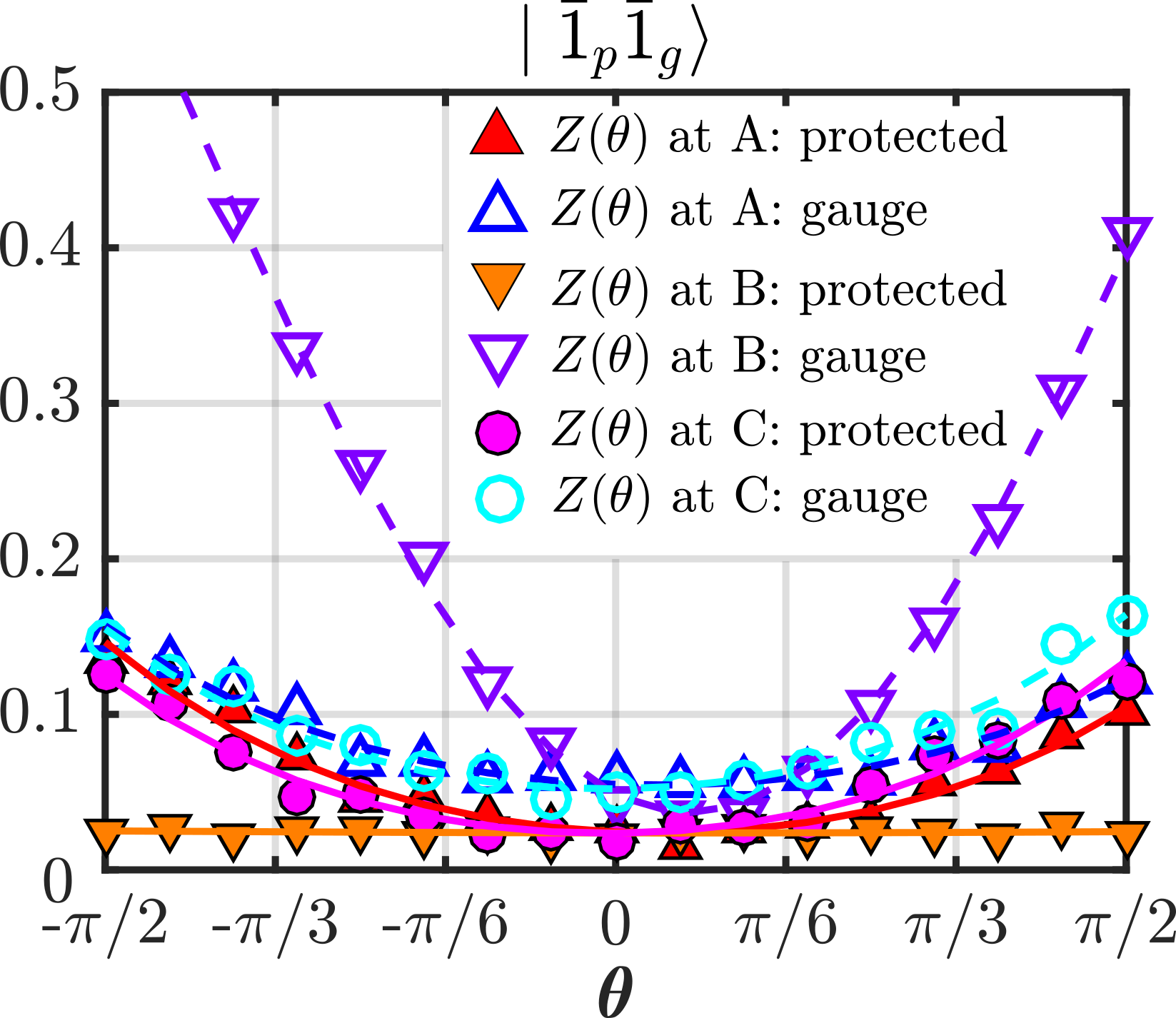}
	}
	\vspace{.05in}
	\subfloat[Acceptance $Y(\theta)\otimes Y(\theta)$ \label{figure:PreH_Error-Accept}]{%
		\includegraphics[width=1.6in]{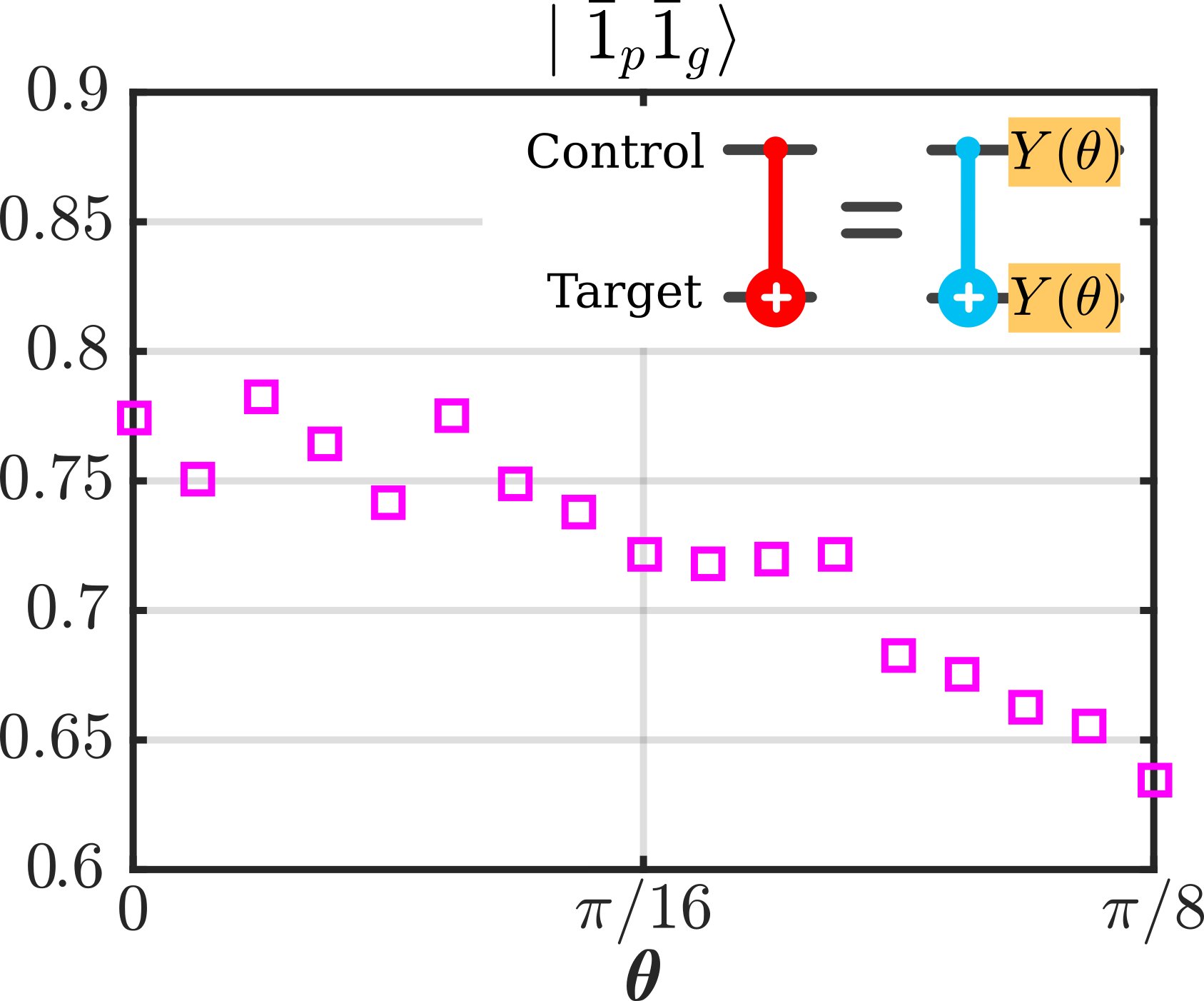}
	}
	\hfill
	\subfloat[Error $Y(\theta)\otimes Y(\theta)$ \label{figure:PreH_Error-probability}]{%
		\includegraphics[width=1.6in]{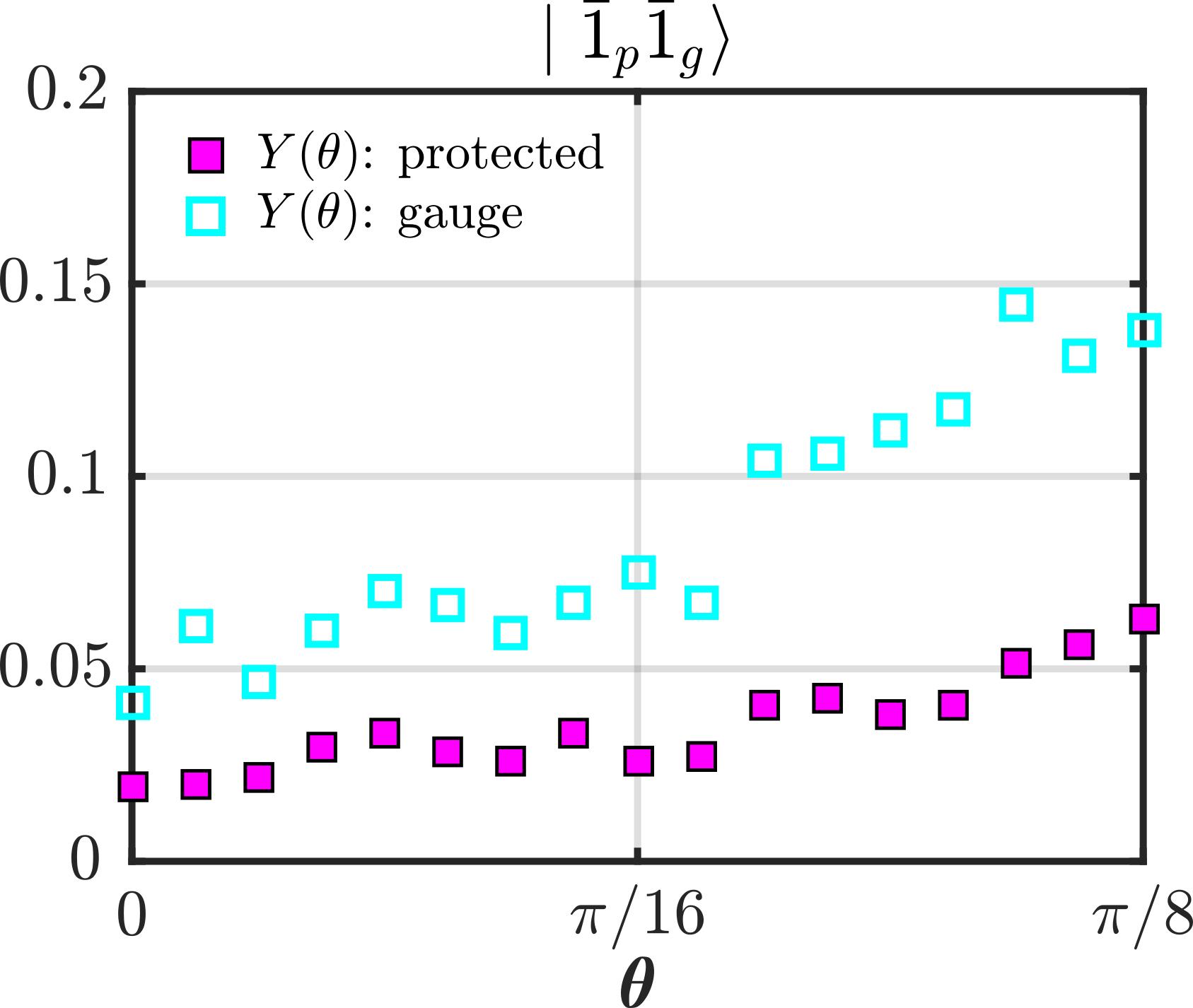}
	}
	\caption{(color online) (a) Acceptance and (b) error probability of logical states with phase error $Z(\theta)$ inserted on the syndrome qubit at various sites. We fit the data to Eq.~\ref{eq:pacceptform} and \ref{eq:logerrform} with an additional systematic offset parameter $\delta$ that is added to $\theta$, see Supplemental Material. (c) Acceptance and (d) error probability of logical states with error $Y(\theta)$ inserted on the control and syndrome qubit after each CNOT gate. \label{figure:ZthetaError}}
\end{figure}

Phase errors propagate from target to control through a CNOT gate, hence a $Z$-error at locations A, B, or C appears as an $X$-error on \{D$_2$,D$_3$,D$_4$\},  \{D$_3$,D$_4$\}, or \{D$_4$\}, respectively. As we increase the error parameter $\theta$, the acceptance probability decreases for locations A and C but remains constant for location B (see Fig.~\ref{figure:ZthetaError} (a)). Fig.~\ref{figure:ZthetaError} (b) plots the state preparation errors as a function of $\theta$. As we increase $\theta$ at location B, error on the protected qubit remains constant, while the error on the gauge qubit increases. For errors inserted at locations A and C, the gauge qubit error is always larger than protected qubit error.

Although a distance two code can only detect one error on the data qubits, correlated two-qubit gate errors are also detectable because the circuit is fault-tolerant by construction. In particular, the two-qubit gates never act directly on pairs of data qubits, so two-qubit gate errors can only affect one data qubit at a time. To mimic this correlated error, we simultaneously introduce $Y(\theta)$ errors on the control and target qubits after each CNOT gate. Similar to single-qubit error insertion, the acceptance probability decreases as a function of $\theta$ (see Fig.~\ref{figure:ZthetaError} (c)) and lower errors are observed on the protected qubit versus the gauge qubit (see Fig.~\ref{figure:ZthetaError} (d)). 

To understand the functional form of the error insertion data, we modeled error insertion in the ideal state preparation circuit followed by asymmetric readout errors with the same readout parameters for each qubit. For each error location, we find the acceptance probability and conditional logical error probabilities as a function of the error's angle $\theta$ and the readout parameters $p_0=P(0|1)$ and $p_1=P(1|0)$. For single $Z(\theta)$ error insertion at location $j$, the acceptance probability has the form
\begin{equation}
P_j(\mathrm{accept}) = a_j + b_j\cos(\theta),\label{eq:pacceptform}
\end{equation}
and the conditional logical error probabilities on logical qubit $r$ have the form
\begin{equation}
P_j(\bar{X}_{Lr}|\mathrm{accept}) = \frac{c^{(r)}_j+d^{(r)}_j\cos(\theta)}{P_j(\mathrm{accept})}.\label{eq:logerrform}
\end{equation}
Each coefficient is a function of $p_0$ and $p_1$ that is given in the Supplemental Material. The expressions for locations A and C are identical. Likewise, for each location, the expression for combined logical error $\bar{X}_{L1}\bar{X}_{L2}$ on the gauge and protected qubit is the same as the corresponding expression for $\bar{X}_{L1}$ alone.

The dashed curves in Figs.~\ref{figure:ZthetaError} (a) and (b) are fits to the functions given in Eq.~\ref{eq:pacceptform} and \ref{eq:logerrform}, but we include a systematic offset parameter $\delta$ that is added to $\theta$, i.e. we replace $\cos(\theta)$ by $\cos(\theta+\delta)$. The offset $\delta_j$ for each location $j$ is determined from either the acceptance or error data based on which has the greatest curvature. The acceptance probability then has 2 remaining free parameters, $\tilde{a}_j$ and $\tilde{b}_j$. Once these are known, each error probability has 2 remaining free parameters $\tilde{c}^{(r)}_j$ and $\tilde{d}^{(r)}_j$.

This model has the benefit of being simple, but it does not include all of the major error sources and most importantly omits dissipation and systematic phase errors that occur during and between two-qubit gates. We have confirmed that the same general functional form is obtained when dissipation is introduced prior to measurement, which is not surprising, since it can be incorporated into readout error parameters of each qubit.

{\it Non-exponential decay under free evolution} -- In this section we study the free evolution of $|\bar{1}\bar{1}\rangle$, post-selected to the codespace of the four-qubit code. Our goal is to observe how decoherence and fixed coupling terms between transmons act on logical states, particularly in the time interval immediately following fault-tolerant state preparation. The experiments are analogous to decay and spin-echo experiments on physical qubits.

Although we are working with encoded states, these decay experiments do not demonstrate a fault-tolerant memory. A fault-tolerant quantum memory would be implemented in this context by repeated syndrome measurements. Repeated syndrome measurements are not feasible in this device due to both technical limitations of measurement durations as well as exponentially decreasing acceptance probability as a function of the number of syndrome measurements. These limitations could be overcome by implementing a quantum error-correcting, rather than error-detecting, code and using a device that is designed for fast, repeated readout.

\begin{figure}[!h]
	\centering
	\includegraphics[width=3.375in]{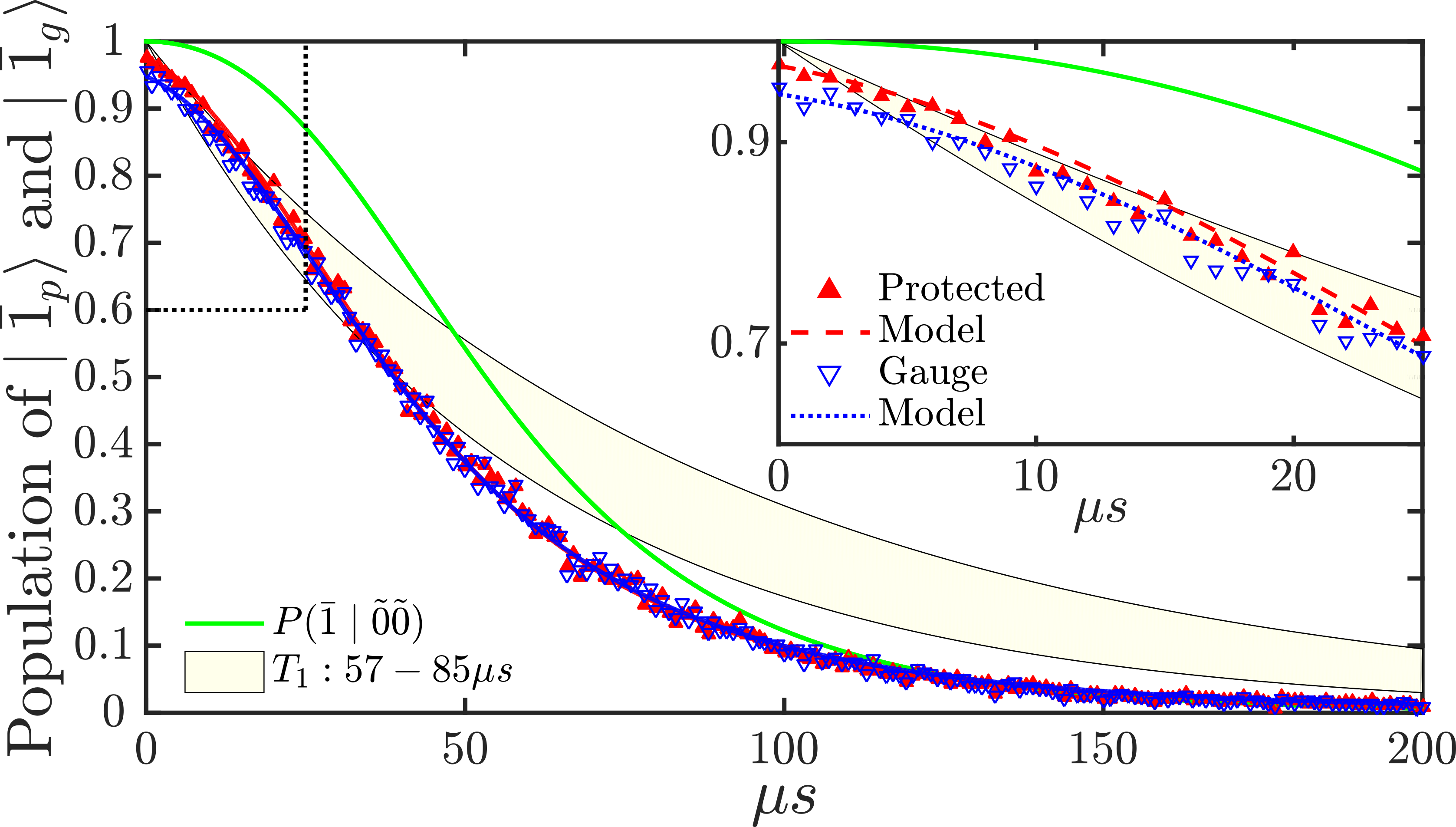}
	\caption{(color online) \textbf{Encoded $\ket{\bar{1}_p \bar{1}_g}$ lifetime}. The ideal curve corresponds to Eq.~\ref{eq:idealdecay}, with $T_1 =76.75\mu$s. Data for the encoded state is plotted with model fits described in the text. Relaxation times of the four data qubits obtained from the model are $T_{1(i)} = \{57,84,85,81\}\mu$s with $i\in{\{1,2,3,4\}}$, which are within one standard deviation of the mean $T_1$ measured for each qubit (see Supplemental Material). The shaded region contains the curves for each qubit from the values of $T_{1(i)}$ obtained from the model fit; $p_0=0.05$ and $p_1=0.015$ are measurement errors from the fit.}
	\label{figure:T11_FPCX}%
\end{figure}

The results for the $|\bar{1}\bar{1}\rangle $ state, shown in Fig.~\ref{figure:T11_FPCX}, have several features that are evidence of short-time protection from local noise. First, the decay is non-exponential, exhibiting a slow initial decay rate that increases with time. The ideal functional form for either encoded qubit is
\begin{equation}\label{eq:idealdecay}
P(\bar{1}|\mathrm{accept}) = (2-2e^{t/T_1}+e^{2t/T_1})^{-1},
\end{equation}
where we have assumed the same $T_1$ for each qubit. A cross-over with the ideal physical decay curve $P(1) = \mathrm{exp}(-t/T_1)$ occurs at $t=T_1\ln 2$, which is on the order of $T_1$. Second, due to fault-tolerant state preparation, the initial population is greater for the protected than the gauge qubit in the presence of error.

To explain how the observed results for the $|\bar{1}\bar{1}\rangle$ state differ from the ideal form, we construct a simplified model of the logical decay incorporating errors in the initial state and readout. The initial state is modeled as
\begin{equation}
\rho(0) = \sum_{L_1,L_2,s_z,s_x} p_{L_1,L_2,s_z,s_x}\rho_{L_1,L_2,s_z,s_x},
\end{equation}
which is a 15 parameter mixture of joint eigenstates $\rho_{L_1,L_2,s_z,s_z}$ of $\bar{Z}_{L1}$, $\bar{Z}_{L2}$, $S_x$, and $S_z$. The parameter values are assigned from state tomography data. Each qubit of this state undergoes independent amplitude damping described by the channel
\begin{equation}
{\cal E}_\gamma(\rho) = A_0\rho A_0^\dagger + A_1\rho A_1^\dagger,
\end{equation}
where $A_0=|0\rangle\langle 0|+\sqrt{1-\gamma}|1\rangle\langle 1|$ and $A_1=\sqrt{\gamma}|0\rangle\langle 1|$. Each qubit has a different damping parameter $\gamma=1-e^{-t/T_1}$ given by a value of $T_1$ that is fitted to experimental data. After damping, each qubit is projectively measured in the computational basis. The readout error process is modeled as an asymmetric binary channel with crossover probabilities $P(0|1)$ and $P(1|0)$. The crossover probabilities are assumed to be the same for each qubit and fitted to the experimental data. Finally, the noisy outcomes are post-processed as described earlier.

{\it Conclusion} -- We demonstrate that even in small code lattices, fault-tolerant principles can result in short-time protection from local dissipation, matching and close to outperforming the evolution of the physical qubits. Due to fault-tolerant circuit design, we observed that one of the two encoded logical qubits has significantly reduced conditional logical error. Additionally, we include quantum state tomography data for prepared codewords, study error insertion, and analyze the decay of measured logical observables under free evolution. The latter shows evidence of short-time protection from local dissipation. A composite two-qubit gate, the four-pulse echoed cross resonance gate, compensated for systematic phase errors during state preparation.  This work, which aims at testing the fundamentals of small codes, is conceived in essence as an effort to understand how noise propagates in larger systems. Repeated stabilizer measurements are needed to study time dependence of fault-tolerant storage.

{\it Acknowledgements} -- We acknowledge Baleegh Abdo for providing the JPCs, Markus Brink for device fabrication, Jim Rozen, Jack Rohrs and Oblesh Jinka for help with the cryogenic setup and Easwar Magesan for discussions on tomography analysis. MT, ADC, and JMC acknowledge  support  from  Intelligence  Advanced Research   Projects   Activity   (IARPA)   under   contract W911NF-16-0114. AWC and JMG acknowledge partial support from ARO under contract W911NF-14-1-0124.
\bibstyle{misc}
\bibliography{qc_422}

\clearpage

\onecolumngrid
\vspace{\columnsep}

\beginsupplement
\section{Supplementary material for `Experimental demonstration of fault-tolerant state preparation with superconducting qubits'}

\section{Device parameters}

\begin{table}[h]
	\begin{tabular*}{3.375in}{l|@{\extracolsep{\fill}}*{5}{c}}
		\hline \hline
		Qubit 		           & $\text{D}_1$ 	    & $\text{D}_2$ 	    & $\text{D}_3$ 	    & $\text{D}_4$ 	    & $\text{S}_1$	\\ \hline
		$\omega_{01}/2\pi$  (GHz)&5.3503		&	5.3061		&	5.229		&5.0748		& 5.1203
		\\ \hline
		$T_1$  & $50.4 $  & $70.3$  & $77.7$  & $68.5$   & $68.0$\\
		$\pm$ std ($\mu$s)   & $ \pm$ 4.8  & $\pm$ 9.7  & $\pm$ 9.8 & $\pm$ 12.5   & $\pm$ 9.6
		\\ \hline
		$T_2$  & $67.6 $  & $104.3$  & $59.6$  & $67.4$  &   $61.6$\\
		$\pm$ std ($\mu$s)  & $ \pm$ 9.3  & $\pm$ 23.3  & $\pm$ 7.1  & $\pm$ 17.3  &   $\pm$ 6.9
		\\ \hline
		$\omega_{r}/2\pi$  (GHz) &	6.5250		& 6.4760		&	6.5742	& 6.5244		& 6.4295 
		\\ \hline
		$\epsilon_{r}$  &	0.0404		& 0.0246		&	0.0323		& 	0.0321	&  0.0256
		\\
		$p_0 = P(0|1)$ &0.0567		&0.0402				&0.0455			&	0.0573			& 0.0424
		\\
		$p_1 = P(1|0)$ &0.0240		&0.0090				&0.0191			&	0.0069			& 0.0088
		\\ \hline \hline 			
		Indiv. RB & 7.59  	& 5.82 	& 6.06 	& 8.56 	& 6.73	\\
		EPG (1e-4)			& $\pm$ 0.12 & $\pm$ 0.10 & $\pm$ 0.10 & $\pm$ 0.16 & $\pm$ 0.10 \\ \hline
		Simul. RB & 13.29  	& 6.81 	& 8.47 	& 9.18 	& 12.29	\\
		EPG (1e-4)		& $\pm$ 0.25 & $\pm$ 0.11 & $\pm$ 0.13 & $\pm$ 0.14 & $\pm$ 0.26 \\ \hline \hline
		
	\end{tabular*}
	\caption{\label{table:device_parameter} \textbf{Qubit and readout characterization.} Qubit transitions ($\omega_{01}/2\pi$), relaxation times ($T_1$), coherence times ($T_2$), readout resonator frequencies ($\omega_{r}/2\pi$), readout assignment errors ($\epsilon_{r}$) obtained from readout parameters, ($P(0|1)$ and $P(1|0)$), and individual and simultaneous single-qubit randomized benchmarking error per gate (EPG) results. Single-qubit gates are all 85 ns long. Anharmonicities of all qubits are around 330 MHz. }	

\end{table}

\begin{table}[h]
	\begin{tabular*}{3.375in}{l|@{\extracolsep{\fill}}*{5}{c}}
		\hline \hline
		& $\text{D}_1$ 	    & $\text{D}_2$ 	    & $\text{D}_3$ 	    & $\text{D}_4$ 	    & $\text{S}_1$	\\ \hline
		$\text{D}_1$ & -& -49& $\epsilon_{zz}$& $\epsilon_{zz}$& -95\\ %
		$\text{D}_2$ & -50& -& $\epsilon_{zz}$& $\epsilon_{zz}$& -29\\ %
		$\text{D}_3$ & $\epsilon_{zz}$& $\epsilon_{zz}$& -& -77& -25\\ %
		$\text{D}_4$ & $\epsilon_{zz}$& $\epsilon_{zz}$& -77& -& -43\\ %
		$\text{S}_1$ & -94& -31& -25& -42& -\\ \hline \hline	          
	\end{tabular*}
	\caption{\label{table:staticZZ} \textbf{Static \textit{ZZ} strength (kHz).}  The strength is measured between each pair of qubits by running a pi-Hahn echo experiment, which is a Hahn echo experiment for each initial state of the spectator qubits in the computational basis, but with a varying angle for the final rotation in the Hahn echo sequence. Here $\epsilon_{zz} < 5$kHz.}
\end{table}

\section{Four qubit code}

\begin{figure}[h]
	\centering
	\begin{tikzpicture}[scale=1.000000,x=1pt,y=1pt]
\draw[] (30.000000,-20.000000) node[] {$|L_1,L_2,s_z,s_x\rangle=|\bar{0}\bar{0},s_zs_x=00\rangle\longmapsto\frac{|0000\rangle+|1111\rangle}{\sqrt{2}}$};
\filldraw[color=white] (0.000000, -7.500000) rectangle (66.000000, 52.500000);
\draw[color=black] (0.000000,45.000000) -- (66.000000,45.000000);
\draw[color=black] (0.000000,45.000000) node[left] {$s_x$};
\draw[color=black] (0.000000,30.000000) -- (66.000000,30.000000);
\draw[color=black] (0.000000,30.000000) node[left] {$L_2$};
\draw[color=black] (0.000000,15.000000) -- (66.000000,15.000000);
\draw[color=black] (0.000000,15.000000) node[left] {$L_1$};
\draw[color=black] (0.000000,0.000000) -- (66.000000,0.000000);
\draw[color=black] (0.000000,0.000000) node[left] {$s_z$};
\begin{scope}
\draw[fill=white] (12.000000, -0.000000) +(-45.000000:8.485281pt and 8.485281pt) -- +(45.000000:8.485281pt and 8.485281pt) -- +(135.000000:8.485281pt and 8.485281pt) -- +(225.000000:8.485281pt and 8.485281pt) -- cycle;
\clip (12.000000, -0.000000) +(-45.000000:8.485281pt and 8.485281pt) -- +(45.000000:8.485281pt and 8.485281pt) -- +(135.000000:8.485281pt and 8.485281pt) -- +(225.000000:8.485281pt and 8.485281pt) -- cycle;
\draw (12.000000, -0.000000) node {$H$};
\end{scope}
\draw (33.000000,15.000000) -- (33.000000,0.000000);
\filldraw (33.000000, 0.000000) circle(1.500000pt);
\begin{scope}
\draw[fill=white] (33.000000, 15.000000) circle(3.000000pt);
\clip (33.000000, 15.000000) circle(3.000000pt);
\draw (30.000000, 15.000000) -- (36.000000, 15.000000);
\draw (33.000000, 12.000000) -- (33.000000, 18.000000);
\end{scope}
\draw (33.000000,45.000000) -- (33.000000,30.000000);
\filldraw (33.000000, 30.000000) circle(1.500000pt);
\begin{scope}
\draw[fill=white] (33.000000, 45.000000) circle(3.000000pt);
\clip (33.000000, 45.000000) circle(3.000000pt);
\draw (30.000000, 45.000000) -- (36.000000, 45.000000);
\draw (33.000000, 42.000000) -- (33.000000, 48.000000);
\end{scope}
\draw (51.000000,30.000000) -- (51.000000,0.000000);
\filldraw (51.000000, 0.000000) circle(1.500000pt);
\begin{scope}
\draw[fill=white] (51.000000, 30.000000) circle(3.000000pt);
\clip (51.000000, 30.000000) circle(3.000000pt);
\draw (48.000000, 30.000000) -- (54.000000, 30.000000);
\draw (51.000000, 27.000000) -- (51.000000, 33.000000);
\end{scope}
\draw (57.000000,45.000000) -- (57.000000,15.000000);
\filldraw (57.000000, 15.000000) circle(1.500000pt);
\begin{scope}
\draw[fill=white] (57.000000, 45.000000) circle(3.000000pt);
\clip (57.000000, 45.000000) circle(3.000000pt);
\draw (54.000000, 45.000000) -- (60.000000, 45.000000);
\draw (57.000000, 42.000000) -- (57.000000, 48.000000);
\end{scope}
\end{tikzpicture}
	\caption{Circuit for encoding logical qubits $L_1$ and $L_2$ into the four-qubit code with syndrome bit values $s_z$ and $s_x$. This encoder is used implicitly in our state tomography analysis, but is not physically implemented. \label{figure:encoder}}
\end{figure}

\section{Fault-tolerant state preparation}

\begin{figure}[hb]
	\subfloat[$|\bar{1}_p\bar{1}_g\rangle$ in $Z$-basis]{%
		\includegraphics[width=3.375in]{delta_L11_FPCX_magnitude.png}
	}
	\subfloat[$|\bar{+}_g\bar{+}_p\rangle$ in the $X$-basis]{%
		\includegraphics[width=3.375in]{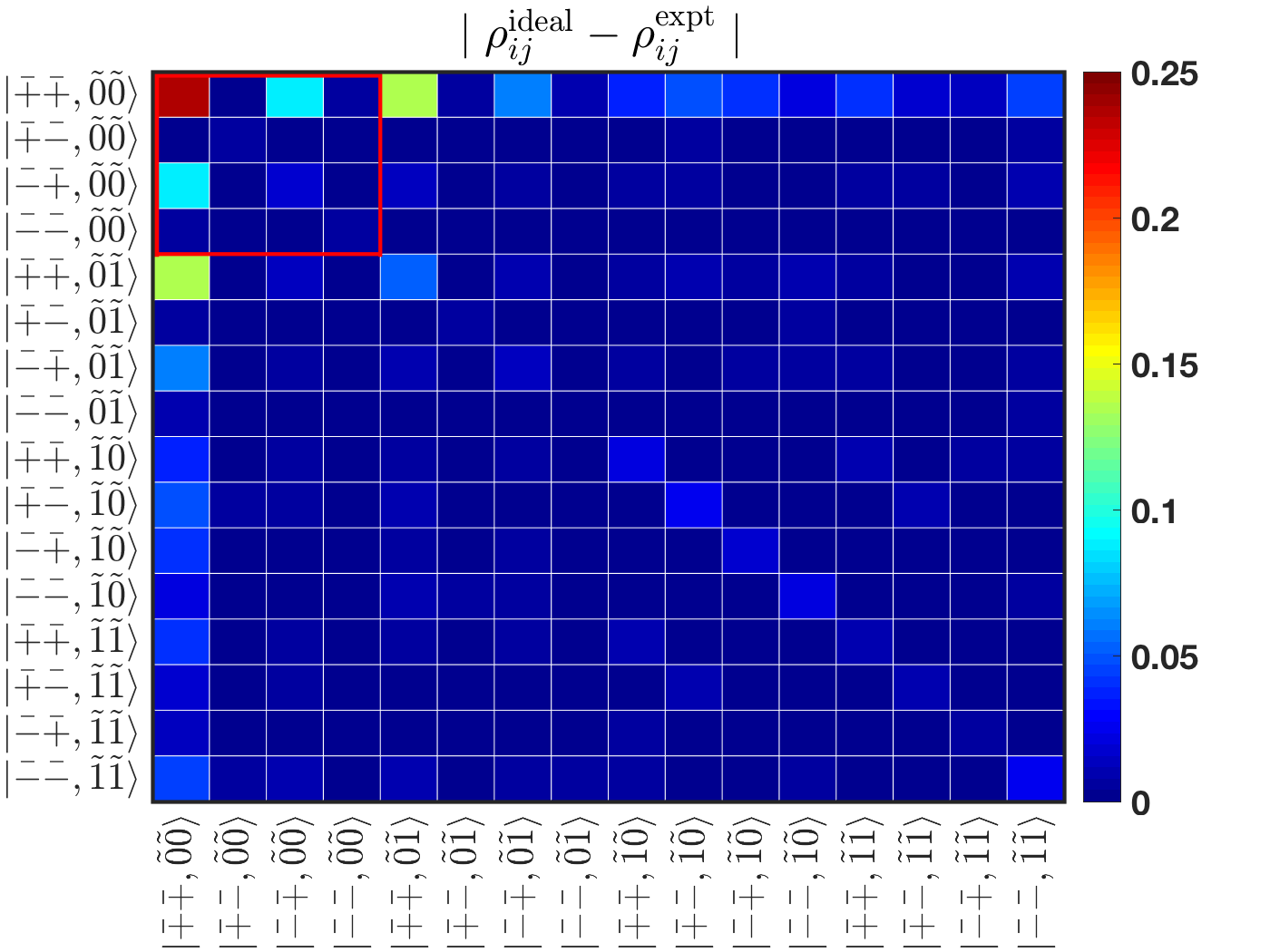}
	}
	\hfill
	\subfloat[$|\bar{1}_p\bar{1}_g\rangle$ in $Z$-basis with TPCX]{%
		\includegraphics[width=3.375in]{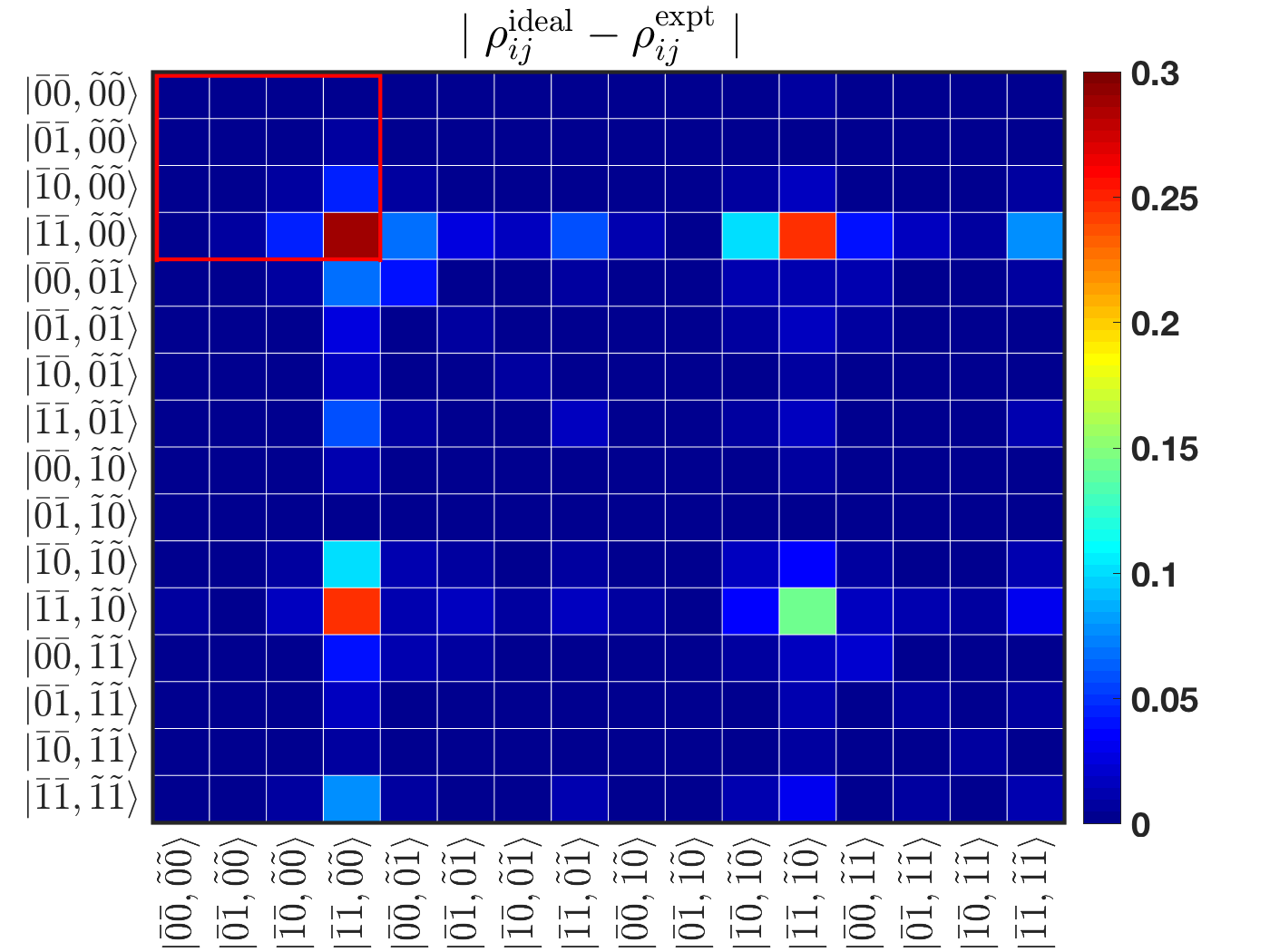}
	}
	\subfloat[$|\bar{+}_g\bar{+}_p\rangle$ in the $X$-basis with TPCX]{%
		\includegraphics[width=3.375in]{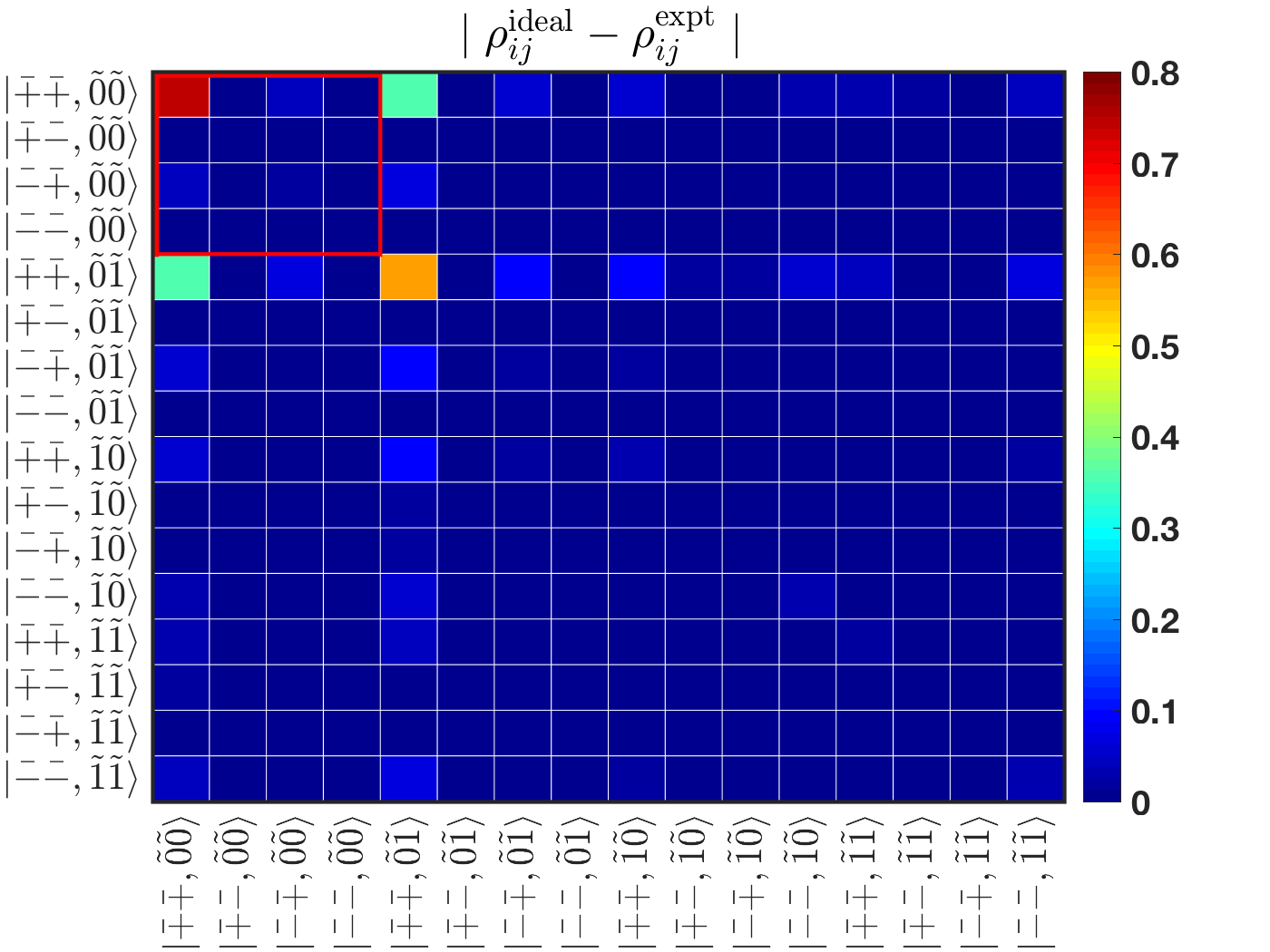}
	}		
	\caption{\label{fig:reconstructed_states}\textbf{Absolute differences between the actual and ideal matrix elements of the reconstructed states.}  The four-qubit Hilbert space is spanned by $16$ states $\{|L_1L_2,s_zs_x\rangle\}$ where $L_1$ and $L_2$ run over the four states of the logical qubits in (a) $Z$-basis and (b) $X$-basis using the FPCX, and (c) $Z$-basis and (d) $X$-basis using the TPCX (see Fig.~\ref{fig:supstateprep} for the pulse decomposition). Syndrome bits $s_z$ and $s_x$ run over the four possible syndromes and represent the presence of phase-flip and bit-flip errors, respectively. The encoding circuit in Fig.~\ref{figure:encoder} makes the logical to physical mapping explicit.}
	
\end{figure}

Fig.~\ref{fig:reconstructed_states} plots absolute differences between the measured and ideal matrix elements of the reconstructed states for $|\bar{1}_p\bar{1}_g\rangle$ and $|\bar{+}_g\bar{+}_p\rangle$ states in $Z$ and  $X$-basis, respectively; we ran state preparation circuits using two different CNOT sequences. Gate decomposition of the two CNOT sequences, the two-pulse CNOT gate (TPCX) and four-pulse CNOT gate (FPCX), are shown in Fig.~\ref{fig:supstateprep}. Two-qubit error per gate results of the two sequences are compared in Table ~\ref{table:RB-result_TPCX}. As discussed in the main letter, first and second order \textit{Z}-terms from the cross-resonance Hamiltonian on control qubit, target qubit, and spectator qubits (SQ) are \textit{ZII, IZI, IIZ, ZZI, ZIZ, IZZ}. While the FPCX sequence echoes all these terms, TPCX does not echo out \textit{IZI, IIZ}, and \textit{IZZ}. 

We observe different sensitivity to systematic phase errors depending on what codeword is prepared using TPCX; Fig.~\ref{fig:supstateprep} shows each preparation circuit. We observed that $|\bar{0}_p\bar{0}_g\rangle$ and hence all states in the $|\bar{\pm}\rangle$ basis were significantly more sensitive to phase error than the other logical basis states $|\bar{0}_p\bar{1}_g\rangle$, $|\bar{1}_p\bar{0}_g\rangle$, and $|\bar{1}_p\bar{1}_g\rangle$. When the $|\bar{+}_g\bar{+}_p\rangle$ state is prepared using TPCX, we observe low acceptance probability (see Table ~\ref{table:AcceptanceAndError_TPCX}), and the dominant state observed from the reconstructed state is the $|\bar{+}_g\bar{+}_p,\tilde{0}\tilde{1}\rangle$ (see Fig. ~\ref{fig:reconstructed_states}). We hypothesize that the largest errors are due to drive-activated phase shifts (AC Stark shifts) during the two-qubit gates, which may have comparable magnitudes $\theta$ and the same sign on each data qubit. The relative phase accumulates as $4\theta$ on $|\bar{0}_p\bar{0}_g\rangle$, but the error effectively cancels on the other states. The FPCX sequence effectively cancels this error.

\begin{table}[h]
	\begin{tabular*}{3.375in}{l|@{\extracolsep{\fill}}*{4}{c}}
		\hline\hline
		&Prepare	&$\:$Accept$\:$	&$\epsilon_p$	&$\epsilon_g$ \\ \hline
		\multirow{2}{*}{$\:$TPCX$\:$}	&$\ket{\bar{1}_p \bar{1}_g}$   				&0.7206	&$\:$0.0035$\:$  	&$\:$0.0137$\:$ \\
		&$\:$$\ket{\bar{+}_g \bar{+}_p}$$\:$  &0.2767  &0.0225 				&0.0587 \\ 
		\hline			
		\multirow{2}{*}{$\:$FPCX$\:$}	&$\ket{\bar{1}_p \bar{1}_g}$   				&0.7853	&$\:$0.0067$\:$  	&$\:$0.0257$\:$ \\
		&$\:$$\ket{\bar{+}_g \bar{+}_p}$$\:$  &0.7897  &0.0134 				&0.0268 \\ 
		\hline\hline
		
	\end{tabular*}
	\caption{\label{table:AcceptanceAndError_TPCX} \textbf{Initial state characterization}. Acceptance probability, error on protected qubit state preparation ($\epsilon_p$), and error on gauge qubit state preparation ($\epsilon_g$) of initial state preparation given that the state is in the codespace, $\tilde{0}\tilde{0}$, using two different CNOT gate sequence.\\ \\}
	
	\begin{tabular*}{0.9\textwidth}{l|@{\extracolsep{\fill}}*{4}{c}}
		
		\hline\hline
		&$CR_1$ 	   &$CR_2$ 		&$CR_3$ 		&$CR_4$ 
		\\ \hline
		TPCX EPG		& $0.0251\pm 0.0010$  	&$0.0199\pm 0.0005$ 	&$0.0170\pm 0.0006$ 	&$0.0169\pm 0.0005$ \\
		(gate length, ns) 	&(435) 		&(475)	&(475)	&(435)
		\\ \hline
		FPCX EPG		& $0.0380\pm 0.0013$  		&$0.0451\pm 0.0015$ 	&$0.0330\pm 0.0012$ 	&$0.0282\pm 0.0010$ \\
		(gate length, ns)	&(780) 		&(780)	&(780)	&(780) \\
		\hline \hline			
		
	\end{tabular*}
	\caption{\label{table:RB-result_TPCX} \textbf{Two-qubit error per gate (EPG).} Randomized benchmarking results and total two-qubit gate times used in TPCX and FPCX.}
\end{table}

\section{Error insertion}

The fitting parameters for Fig. 4 of the main text are as follows. The offset parameters $\delta_\textrm{A}=-0.1369$ and $\delta_\textrm{C}=0.0278$ are determined from the acceptance probability, and the parameter $\delta_\textrm{B}=-0.2291$ is fitted from the conditional logical error probability. The acceptance probability has parameters $\tilde{a}_\textrm{A}=0.5044$ and $\tilde{b}_\textrm{A}=0.2632$ at location A, $\tilde{a}_\textrm{B}=0.7614$ and $\tilde{b}_\textrm{B}=0.0059$ at location B, and $\tilde{a}_\textrm{C}=0.4983$ and $\tilde{b}_\textrm{C}=0.2708$ at location C. The conditional logical error for the protected qubit has parameters $\tilde{c}^{(1)}_\textrm{A}=0.0626$ and $\tilde{d}^{(1)}_\textrm{A}=-0.0444$ at location A, $\tilde{c}^{(1)}_\textrm{B}=0.0189$ and $\tilde{d}^{(1)}_\textrm{B}=-0.0006$ at location B, and $\tilde{c}^{(1)}_\textrm{C}=0.0646$ and $\tilde{d}^{(1)}_\textrm{C}=-0.0466$ at location C. The conditional logical error for the gauge qubit has parameters $\tilde{c}^{(2)}_\textrm{A}=0.0697$ and $\tilde{d}^{(2)}_\textrm{A}=-0.0279$ at location A, $\tilde{c}^{(2)}_\textrm{B}=0.3847$ and $\tilde{d}^{(2)}_\textrm{B}=-0.3573$ at location B, and $\tilde{c}^{(2)}_\textrm{C}=0.0795$ and $\tilde{d}^{(2)}_\textrm{C}=-0.0395$ at location C. The tilde denotes parameters from fitting data to Eq. 9 and 10 of the main text.

For the simplified error insertion model described in the main text, the acceptance probability coefficients are
\begin{align*}
a_\textrm{A} & = \frac{1}{2}\left( 1 + (p_0-p_1)^2 (3 + 4p_0^2 - 6p_1 + 4p_1^2 + p_0(4p_1-6)) \right), \\
b_\textrm{A} & = \frac{1}{2} (p_0+p_1-1)^2 (1 + 4p_0^2 - 2p_1 + 4p_1^2 - 2p_0(2p_1+1)), \\
a_\textrm{B} & = 1 + 2 (p_0(p_0 - 1) + p_1(p_1-1)) (1 + p_0(p_0-1) + p_1(p_1-1)), \\
b_\textrm{B} & = (p_0 - p_1)^2 (p_0 + p_1 - 1)^2.
\end{align*}
The coefficients in the conditional logical error probability are
\begin{align*}
c^{(1)}_\textrm{A} & = \frac{1}{4}(2p_0^4 + p_1 + 3p_0^2p_1+p_1^3(2p_1-3) -p_0^3(2p_1+3)+p_0(1+p_1(-2+(3-2p_1)p_1))), \\
d^{(1)}_\textrm{A} & = \frac{1}{4}(p_0+p_1-1)^2(2p_0^2+p_1(2p_1-1)-p_0(2p_1+1)), \\
c^{(1)}_\textrm{B} & = \frac{1}{2}(p_0(p_0-1)+p_1(p_1-1))^2, \\
d^{(1)}_\textrm{B} & = \frac{1}{2}(p_0-p_1)^2(p_0+p_1-1)^2, \\
c^{(2)}_\textrm{B} & = \frac{1}{2}(p_0^2+(p_1-1)^2)((p_0-1)^2+p_1^2), \\
d^{(2)}_\textrm{B} & = \frac{1}{2}((p_0-p_1)^2-1)(p_0+p_1-1)^2.
\end{align*}
The $L_2$ (gauge) logical error probability at location A is the same as the $L_1$ (protected) logical error. We find that $\tilde{p}_0=0.108$ and $\tilde{p}_1=0.043$ minimize the sum of the absolute differences between the model and fitted parameters.

\section{Encoded $|\bar{+}\bar{+}\rangle$ lifetime}

\begin{figure}[!h]
	\centering
	\includegraphics[width=3.375in]{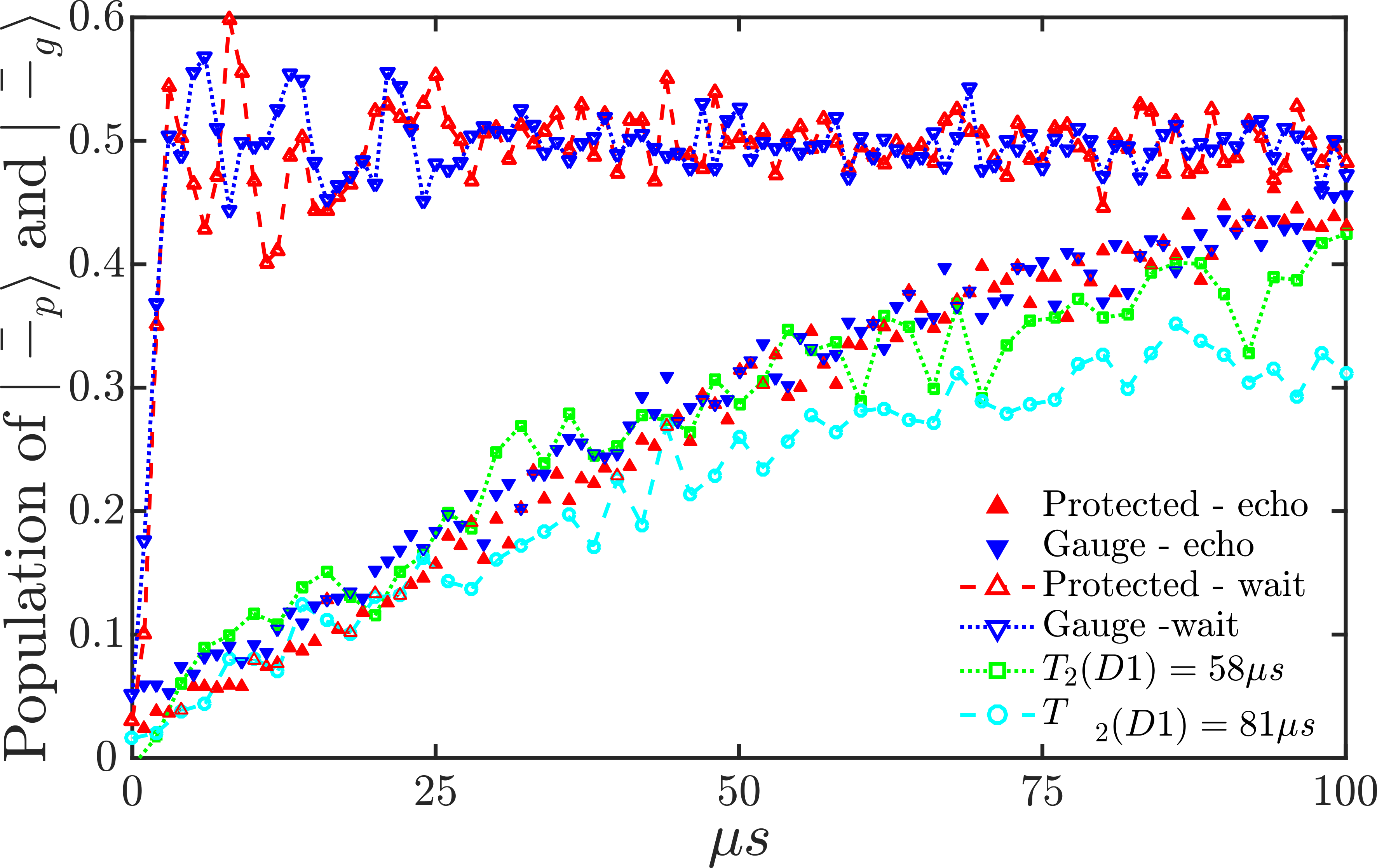}
	\caption{\textbf{Encoded $\ket{\bar{+}_g \bar{+}_p}$ lifetime.} Without echo pulses, the codeword  decays to $1/2$ in about $4 \mu$s and oscillates for about $20 \mu$s. Using an echo sequence leads to comparable decay rates to that of the physical qubits.}
	\label{figure:TPP_FPCX}%
\end{figure}

\begin{figure}[p]
	\centering 
	\includegraphics[width=7in]{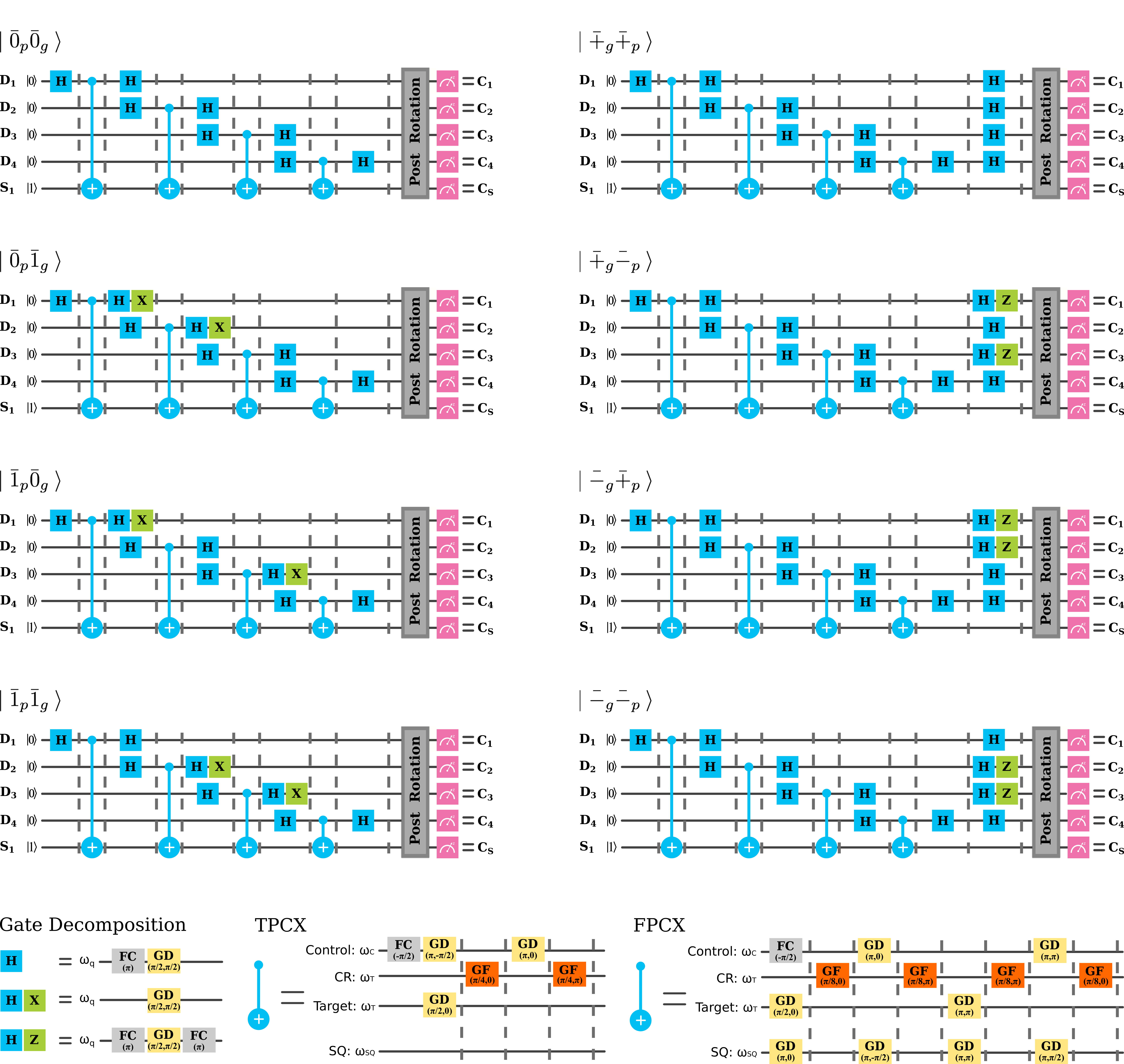}
	\caption{\textbf{Preparation circuits.} These circuits prepare each of the logical basis states in the standard and $|\bar{\pm}\rangle$ basis. One important difference in these circuits is the placement of the Pauli operators that map from $|\bar{0}_p\bar{0}_g\rangle$ and $|\bar{+}_g\bar{+}_p\rangle$ to the other states, as this affects how phase accumulates while qubits are idle.\label{fig:supstateprep}}
\end{figure}

The experimental results for the $|\bar{+}\bar{+}\rangle $ state are shown in Fig.~\ref{figure:TPP_FPCX}. There is rapid decay and coherent oscillation for both encoded qubits. This decay can be slowed by using echo sequences. The rapid decay and coherent oscillation is consistent with free evolution under the static ZZ terms in the Hamiltonian. Taking a ZZ strength $\eta$ of order $50$ kHz gives timescales $t_\pi = \pi/|\eta| = (2\times -50 \mathrm{kHz})^{-1} = 10\mu$s, which is consistent with the loss of coherence in Fig.~\ref{figure:TPP_FPCX}.

\end{document}